\documentclass[12pt]{iopart}

\usepackage{graphicx}
\usepackage{dcolumn}
\usepackage{bm}
\usepackage{hyperref}

\usepackage[english]{babel}
\usepackage[utf8x]{inputenc}
\usepackage{iopams}

\expandafter\let\csname equation*\endcsname\relax 
\expandafter\let\csname endequation*\endcsname\relax
\usepackage{amsmath}

\usepackage{graphicx}
\usepackage[colorinlistoftodos]{todonotes}

\begin{document}

\title{Single photon production by rephased amplified spontaneous emission}
\author{R N Stevenson$^1$, M R Hush$^2$, A R R Carvalho$^{1,3}$, S E Beavan$^4$, M J Sellars$^3$ and J J Hope$^1$}
\address{1 Department of Quantum Science, Research School of Physics and Engineering, The Australian National University, ACT 0200, Australia}
\address{2 School of Physics and Astronomy, University of Nottingham, Nottingham, NG7 2RD, United Kingdom}
\address{3 ARC Centre for Quantum Computation and Communication Technology, The Australian National University, ACT 0200, Australia }
\address{4 Center for NanoScience and Fakultät für Physik, Ludwig-Maximilians-Universität, Geschwister-Scholl-Platz 1, 80539 München, Germany}

\begin{abstract}
The production of single photons using rephased amplified spontaneous emission is examined. This process produces single photons on demand with high efficiency by detecting the spontaneous emission from an atomic ensemble, then applying a population-inverting pulse to rephase the ensemble and produce a photon echo of the spontaneous emission events. The theoretical limits on the efficiency of the production are determined for several variants of the scheme. For an ensemble of uniform optical density, generating the initial spontaneous emission and its echo using transitions of different strengths is shown to produce single photons at 70\% efficiency, limited by reabsorption. Tailoring the spatial and spectral density of the atomic ensemble is then shown to prevent reabsorption of the rephased photon, resulting in emission efficiency near unity.
\end{abstract}

\submitto{\NJP}

\maketitle

Single photons have emerged as an integral parts of a range of quantum protocols, from computing \cite{Browne:2005} to random number generation \cite{Rarity:1994} and digital security \cite{Bennett:1984}. Single photon states, and photonic Fock states with well defined numbers of photons, are in practice difficult to produce.  Desirable properties for a single photon source are that the emitted photons are in a tight spatial mode; that they be available on demand, or at least have a heralded arrival; that they be produced with high efficiency; and that they have negligible populations of higher numbers of photons.  A range of sources have been proposed and demonstrated~\cite{Eisaman:2011}, including pair sources, such as through parametric downconversion~\cite{Louisell:1961,Burnham:1970}, single emitter sources, such as single atoms~\cite{McKeever:2004}, and ensemble sources, such as the DLCZ protocol~\cite{Duan:2001}. These ensemble sources in particular allow for the collective effects of multiple emitters to be harnessed, producing light in highly directional spatial modes through mode-matching~\cite{Beavan:2012a}.

\begin{figure}[ht]
\begin{center}
\includegraphics[width=30pc]{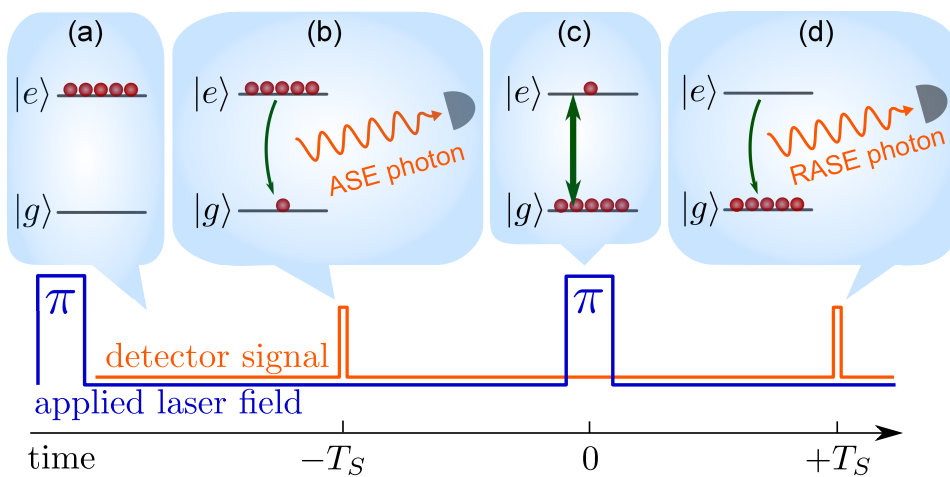}
\end{center}
\caption{\label{fig:RASE}
The RASE protocol and corresponding detector signals. The ensemble is initially prepared in the ground state (a) A $\pi$-pulse puts the ensemble in the excited state. (b) A spontaneous emission event is detected, leaving a single collective de-excitation in the ensemble. (c) A second $\pi$-pulse inverts the population, leaving the ensemble with a single collective excitation. (d) The excitation rephases and the ensemble emits an echo of the first emission.
}
\end{figure}

A possible strategy to generate photons using atomic ensembles is through the process of Rephased Amplified Spontaneous Emission (RASE)~\cite{Ledingham:2010,Beavan:2012a,Ledingham:2012}. In this process (see Fig.\ref{fig:RASE}), the ensemble is initially prepared in the excited state from which it can spontaneously emit.
Direct photodetections of these collective spontaneous emissions, which will be referred to in this paper as the Amplified Spontaneous Emission (ASE), project the system into a state of collective de-excitations distributed over the ensemble.
These de-excitations dephase through inhomogeneous broadening. The population is then inverted by a $\pi$-pulse. This turns the collective de-excitations into collective excitations, which rephase due to an effective time-reversal of the inhomogeneity. When the ensemble is larger in at least one dimension than the wavelength of the transition, then the rephased excitations are emitted in a highly directional spatial mode satisfying the phase matching condition.

In order to build a faithful single photon source using RASE, one needs a single ASE event, or, at least, that multiple detection events are well separated in time. Moreover, for the source to be efficient, one needs a high probability of rephasing the emitted photon. In this paper, we model the quantum dynamics of the RASE process conditioned on the detection of emitted photons to find the regime of operation where these conditions are satisfied. In particular, we show that to avoid multiple photon events one should work in the low optical depth regime, while rephasing efficiency approaches unity with increasing optical depth. We also show that the problem of incompatibility of these two regimes can be circumvented by an appropriate tailoring of the spectral density of the ensemble. This strategy allows, in principle, a true single photon source based on RASE to reach unity efficiency.

The paper is organised as follows: In Section \ref{sec:model} we describe the model of ASE from an ensemble when it is monitored by a photon counting detector, and the RASE of a single collective excitation. In section \ref{sec:analysis} we find the effectiveness of RASE for producing single photons, taking into account the balance between improving the efficiency of emission of rephased photons while minimising the chance of multi-photon events. In Section~\ref{sec:tailored}, we describe a modification of the RASE scheme that uses different atom-light coupling for the ASE and RASE emissions and shaped density profiles of the ensemble to increase the effectiveness of the production of single photons.

\section{Modelling ASE and RASE}
\label{sec:model}

The two stages of the RASE protocol are modelled separately. The first stage models ASE from an excited state ensemble up until the ensemble is inverted (Fig.\ref{fig:RASE}- a and b). For this stage the ensemble is modelled as an open quantum system coupled to a bath of optical modes that are monitored by a photon-counting detector. The second stage models RASE from this ensemble from after the ensemble is inverted (Fig.\ref{fig:RASE}- c and d). In this stage the ensemble and optical modes are modelled together as a closed quantum system.

The atomic ensemble consists of a large number of individual atoms fixed in space with varying micro-environments that change their resonant frequency significantly, even over length scales smaller than a wavelength.  We approximate the ensemble as a continuous system in position ($z$) and detuning ($\Delta$) of the atomic resonance from a central frequency. When the atomic ensemble is significantly larger than the wavelength of the light, phase-matching conditions are only satisfied if the RASE photons emit in a tight spatial mode around the seed ASE photon~\cite{Beavan:2012a}. For this reason we can treat the system as being spatially one-dimensional. The continuous operators for this system have the following commutation relations: 

\begin{align}
[\hat \sigma_+(z,\Delta, \hat \sigma_-(z',\Delta')] &= \hat \sigma_3(z, \Delta)\delta(z-z') \delta(\Delta - \Delta')\\
[\hat \sigma_+(z,\Delta), \hat \sigma_3(z',\Delta')] &= -\hat \sigma_+(z, \Delta)\delta(z-z') \delta(\Delta - \Delta')\\
[\hat \sigma_-(z,\Delta), \hat \sigma_3(z',\Delta')] &= \hat \sigma_-(z, \Delta)\delta(z-z') \delta(\Delta - \Delta').
\end{align}
Here $\delta$ is the Dirac delta function. The operators $\hat \sigma_\pm(z,\Delta)$ are the raising (+) and lowering (-) operator for atoms at position $z$ with detuning $\Delta)$. The operator $\hat \sigma_3 (z, \Delta)$ is the population density difference operator. This operator acts on a state with all atoms in the excited state $|\{e\}\rangle$ or all atoms in the ground state $|\{g\}\rangle$ in the following way:

\begin{align}
\hat \sigma_3(z,\Delta) |\{e\}\rangle &= \rho(z, \Delta) |\{e\}\rangle\\
\hat \sigma_3(z,\Delta) |\{g\}\rangle &= -\rho(z, \Delta) |\{g\}\rangle,
\end{align} 
where the function $\rho(z,\Delta)$ describes the density of atoms at position $z$ with detuning $\Delta$. This combined spatial and spectral density of active atoms characterises the main difference between different types of atomic ensembles.

\begin{figure}[ht]
\begin{center}
\includegraphics[width=30pc]{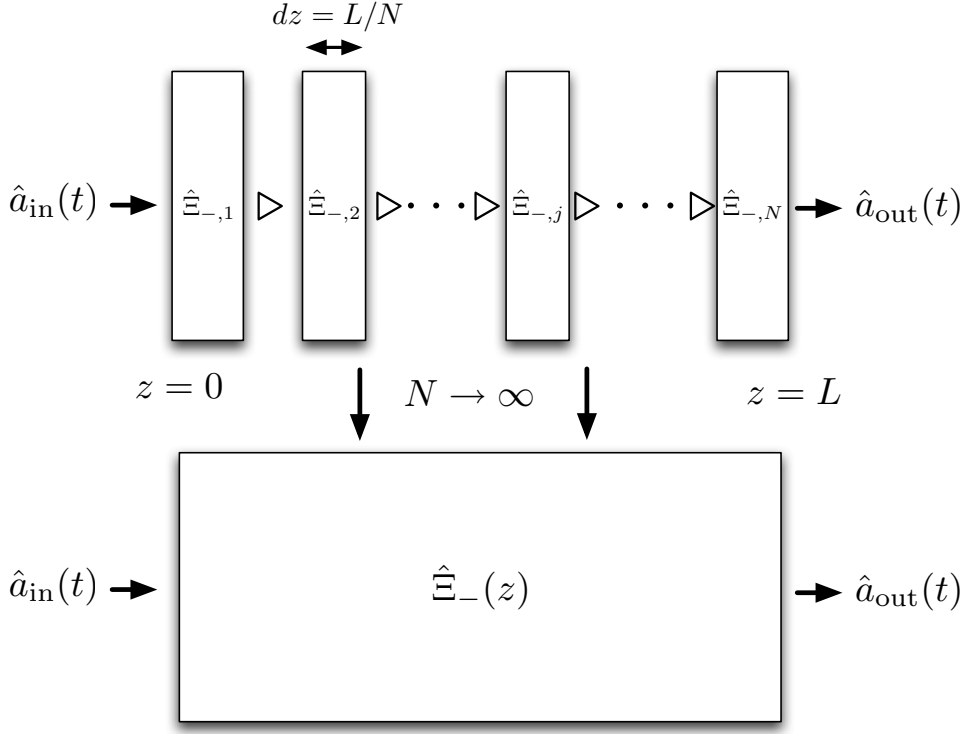}
\end{center}
\caption{\label{fig:model}
The action of the atomic ensemble on the input light ($\hat a_\text{in}(t)$) to give the output light ($\hat a_{out}(t)$) can be found by combining the actions of thin slices of the ensemble using a series product ($\triangleright$).
}
\end{figure}

In both ASE and RASE periods, the dynamics is complicated by the fact that light emitted from one part of the ensemble can interact with other parts of the ensemble before it exits. We model this process using an input-output approach, where we consider the light exiting the $n$-th spatial slice of the ensemble as the input of $n+1$ slice, as shown in Fig.~\ref{fig:model}. This allows us to use the quantum network approach developed by James and Gough~\cite{James:2010}. The slices form a network of elements connected in series or, in the quantum optics language, a set of cascade quantum systems~\cite{Gardiner:1993, Carmichael:1993a}.

In the quantum network theory the properties of a quantum system are described by three operators: the Hamiltonian $\hat H$ that describes the internal evolution of the quantum system; the output operators $\bf \hat L$ that describe the coupling of the internal state of the system to the output channels; and the scattering matrix $\hat S$ that describes the way in which the input channels are mixed to give the output channels. Similar to the approach in~\cite{Hush:2013} describing quantum memories, in our case the atomic ensemble is broken down into slices, with the output of one slice being the input for the next slice. Each slice contains a collection of atoms with different detunings, but the slices are considered so thin that there are no emission then re-absorption events within a single slice. As these atoms are very close to the excited state during ASE the population difference operator is approximately constant: $\hat \sigma_3(z,\Delta) \approx \rho(z,\Delta)$, so these operators behave similarly to harmonic oscillator raising and lowering operators~\cite{Ledingham:2010}. The triplet of operators $G_j=\left(\bf {\hat{S}}_j,{\bf \hat{L}}_j,\hat{H}_j\right)$ describing atom in slice $j$ with a single input and single output channel is~\cite{James:2010}

\begin{align}
G_j = \left( \hat 1,\,\, i \frac{g}{\sqrt c} \hat \Xi_{-,j}, \,\, \hbar\,  \hat\Xi_{3,j}\right),
\end{align}
where $g$ is the coupling between the atom and the vacuum, $c$ is the speed of light, $\hat\Xi_\pm (j) = \int d\Delta \hat\sigma_{\pm,j}(\Delta)$ and $\hat\Xi_{3,j} = \int d\Delta\, \Delta \hat\sigma_{-,j}(\Delta)\hat\sigma_{+,j}(\Delta)$. 

The process by which the output of one system is used as the input for another is called the series product. The series product rule allows one to write the operator describing the compound system in terms of the operators of the individual systems. One spatial dimension, and therefore only one input and output channel, gives a scattering matrix of unity. The series product of two slice with index 1 and 2 is~\cite{James:2010}

\begin{align}
&(\hat 1,\hat L_1, \hat H_1) \triangleright (\hat 1, \hat L_2, \hat H_2)= \left( \hat 1,\,\, \hat L_1 + \hat L_2,\,\, \hbar \, \hat H_1 + \hbar \,  \hat H_2 + \frac{\hbar}{2i} \left(\hat L_2^\dagger \hat L_1 - \hat L_1^\dagger \hat L_2\right)\right).
\end{align}
This series product can be used to combine $N$ slices of our ensemble, as shown in figure~\ref{fig:model}:

\begin{align}
G &= G_1 \triangleright G_2 \triangleright ... \triangleright G_N \nonumber\\
&= \left( \hat 1,\,\, \sum_j \hat L_j, \,\,\hbar \sum_j \hat H_j + \frac{\hbar}{2i}\sum_{k=2}^N \sum_j^{k-1}  \left(\hat L_k^\dagger \hat L_j - \hat L_j^\dagger \hat L_k\right)\right)\nonumber\\
&= \Bigg( \hat 1, \,\,\frac{g}{\sqrt c}\sum_j i  \hat \Xi_{-,j}, \,\hbar \sum_j \hat\Xi_{z,j} + \frac{\hbar}{2i}\frac{g^2}{c}\sum_{k=2}^N \sum_j^{k-1}  \left(\hat  \hat \Xi_{-,k}^\dagger \hat \Xi_{-,j} - \hat \Xi_{-,j}^\dagger \hat \Xi_{-,k}\right)\Bigg).
\end{align}
Taking the limit as the number of slices goes to infinity we get

\begin{align}
(\hat S, \hat L, \hat H) &= \Bigg(\hat 1, i \frac{g}{\sqrt c} \int_0^L dz \, \hat \Xi_-(z),\hbar \int_0^L dz \, \hat \Xi_z (z) \nonumber\\
&\hspace{5mm}+ \frac{\hbar}{2i}\frac{g^2}{c} \int_0^L dz 
\int_0^z dy \left(\hat\Xi_+(z)\hat\Xi_-(y) - \hat\Xi_+(y)\hat\Xi_-(z)\right)\Bigg). 
\label{Eqn:Triplet}
\end{align}
These operators describe the evolution of the atomic ensemble, and the interaction of the ensemble with the bath. This triplet can now be used to find the conditional master equation for the monitored system, which can be simulated using quantum trajectory methods~\cite{Carmichael:1993, Molmer:1993}. Since our monitoring scheme is based on photodetections, the dynamics can be described in terms of quantum jumps. If a photon is detected, the (un-normalised) state of the system changes according to 
\begin{align}
\label{eq:jump}
|\psi^{\prime}\rangle =\hat L|\psi \rangle
\end{align}
where $\hat L$ is the operator describing the effect of the jump in the system.
If no detection events are recorded, the system evolves smoothly under the non-Hermitean Hamiltonian $\hat H_{\rm eff}=\hat H - i \hbar \hat L^\dagger \hat L/2$, and the (un-normalised) state obeys the following dynamics 
\begin{align}
\label{eq:nojump}
\frac{d|\psi(t)\rangle}{dt} = -\frac{i}{\hbar} \hat H |\psi\rangle -\frac{1}{2} \hat L^\dagger \hat L|\psi\rangle.
\end{align}

\subsection{Amplified Spontaneous Emission}
\label{sec:distribution}

We are now in position to analyse the different stages of the RASE protocol. Let us begin by examining the emission process in ASE and what happens when the first emitted photon is detected. Using Eq.~(\ref{eq:jump}) with the $\hat L$ operator in Eq.~(\ref{Eqn:Triplet}), we can write the state of the system following the first detection event as:
\begin{align}
\label{eq:singleEmission}
|\psi\rangle &= \hat L|\{e\}\rangle\nonumber\\ 
&= \int dz \int d\Delta \, \rho(\Delta, z) \hat \sigma_-(z,\Delta) |\{e\}\rangle.
\end{align}
Note that, immediately after the detection, this de-excitation is equally distributed among all atoms. However, as time passes, the system will evolve and the excitation distribution will change. We will assume that no more detection events occur and therefore we can consider only states with a single de-excitation. These states can be written in their general form as:
\begin{align}
|\psi(t)\rangle = \int dz \int d\Delta s(z,\Delta, t) \hat \sigma_-(z,\Delta) |\{e\}\rangle.
\label{eqn:generalASE}
\end{align}
Under the assumption of no further detections, the state will evolve under Eq.~(\ref{eq:nojump}) and we can write the equation of motion for the coefficient $s(z,\Delta,t)$ as
\begin{align}
\frac{d s(z,\Delta,t) }{dt} = \, &i \Delta \, s(z,\Delta,t) - \frac{\pi g^2}{c} (N-1) s(z,\Delta,t)\nonumber\\
&- \frac{2\pi g^2}{c}\int_z^L dy \int d\Delta' s(y, \Delta',t)\rho(y,\Delta').
\label{Eqn:1photon}
\end{align}
Solving this equation gives the unnormalised state of the system conditioned on the absence of further photodetections, and the norm of the state gives the probability that there have been no further emissions:
\begin{align}
P_\text{no jump} (t) = \int dz \int d\Delta |s(z,\Delta, t)|^2 \rho(z,\Delta)
\label{eq:noJump}
\end{align}

It is reasonable to expect that there may be multiple spontaneous emission events within the ASE period.  A general description of two photons would allow the possibility of entangled photons, but when they interact with the ensemble at well-separated times, the combined evolution factorizes into single-photon modes as described by equation (\ref{Eqn:1photon}), where the $m$th photon has the $(N-1)$ factor replaced by $(N-m)$.  This occurs because a short time after the detection of the previous emission, the third term in equation (\ref{Eqn:1photon}) becomes insignificant due to the phase rotation of the collective state, removing any non-trivial possibility of entangling the photons. Operating the single photon source with strong temporal resolution between subsequent photons allows more of the photons produced to be treated as single photons, so we operate in regimes where the above approximation can be used freely.  As the number of atoms in the ensemble is large, we may further assume that $N>>m$, and therefore model all emissions using the same equations as the initial emission.

\begin{figure}[htb]
\begin{center}
\includegraphics[width=17pc]{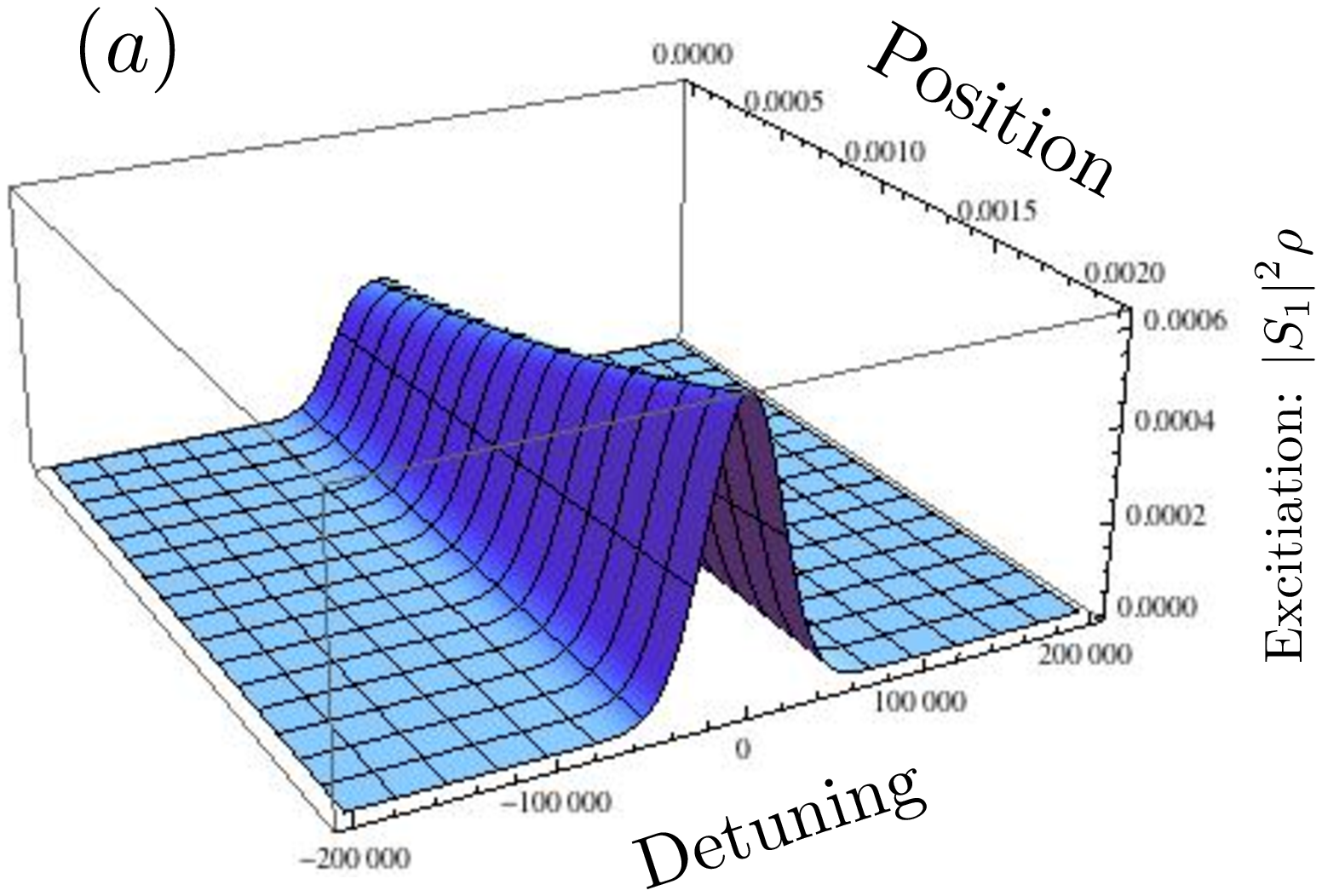}
\includegraphics[width=17pc]{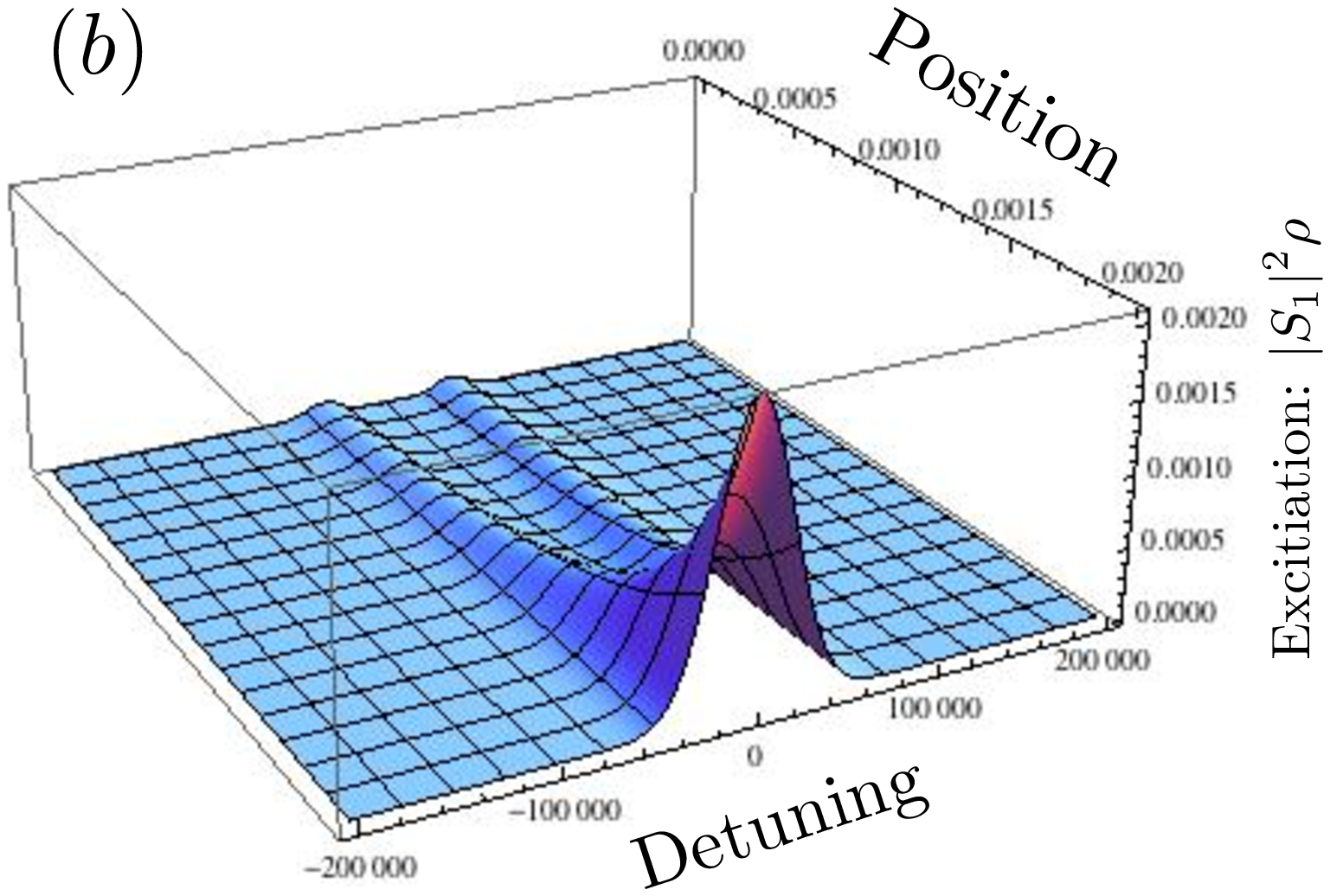}
\end{center}
\caption{\label{fig:shapes} Distribution of ensemble de-excitation at a time $4/Gamma$ after an ASE emission of a single photon. The emission direction is the front of the ensemble, at $z$ = 0.2mm. Figure (a) is calculated for an ensemble with a peak optical depth of 0.75, and the distribution is close to an equal superposition of each atom emitting, mirroring the density of the ensemble, though the front has a slightly higher share of the de-excitation than the back. Figure (b) is calculated for an ensemble with a peak optical depth of 7.5, and the de-excitation is concentrated at the front, though there are ridges of de-excitation extending to the back of the ensemble off-resonant from the central peak. These ridges are due to the ensemble being less dense at these frequencies, so emissions are more likely to traverse the ensemble without stimulating more emissions.}
\end{figure}

The calculation of the photon emissions at higher optical depths shows interesting structure in the spatial and spectral distribution of the collective de-excitation in the ensemble.  Each emission instantaneously produces a uniformly distributed de-excitation, but in the time following the emission from the ensemble, the distribution of the de-excitation changes.  This reshaping reaches a steady state on a timescale inversely proportional to the inhomogeneous linewidth of the ensemble. For a Gaussian distribution with standard deviation $\Gamma$ the shape reaches steady state at around $4 \times 1/\Gamma$.  This is the same timescale found for the photon pulse length in section~\ref{sec:spacing}.  Figure \ref{fig:shapes} shows this stable distribution of the excitation for two different optical depths.

We can understand why this evolution follows the same timescale as the temporal envelope of the photon: after the photon has left the sample, there are no mechanisms to redistribute the population, so all that is left is the phase evolution.  Similarly, the final shape can be understood by considering the fact that photons emitted from the back of the sample have a chance of being reabsorbed at the front, so the higher the optical depth, the more likely it is that a photon that has left the sample came from the front.  We see this clearly in figure \ref{fig:shapes}, and we also see that this effect is more pronounced at resonance, leaving some non-trivial structure at high optical depths. 

\subsection{Rephased Amplified Spontaneous Emission}
\label{sec:RASE}

After ASE has been detected and the ensemble has been inverted, we are interested in modelling the emission of a specific, single collective excitation. As the Hilbert space of a single excitation is small, this can be directly modelled using the Schr\"odinger equation, explicitly including both the atomic ensemble and the light field. The Hamiltonian for the system, with the light in the position basis, is 

\begin{align}
\frac{\hat H}{\hbar} &= \int dz \int d\Delta\, \Delta |e_{z,\Delta}\rangle\langle e_{z,\Delta}| -i \int dz \,c \,\hat a(z)^\dagger \frac{d}{dz} \hat a(z)\nonumber\\
&\hspace{5mm}+\sqrt{2\pi} \int d z \int d \Delta \, g \hat \sigma_+(z,\Delta) \hat a(z) + \text{H.c.} \label{eqn:schrodinger}
\end{align}
where $|e_{z,\Delta}\rangle$ is the state where the ensemble is excited at position $z$ and detuning $\Delta$. The first term is the atomic energy, the second term is the light energy, and the third term is the atom-light interaction.

A general state for a single excitation distributed among the atomic ensemble and the light field is
\begin{align}
|\psi\rangle &= \Bigg(\int dz \int d\Delta s(z,\Delta,t) \hat \sigma_+(z,\Delta) + \int dz \phi(z,t) \hat a^\dagger(z)\Bigg)|\{g\},0\rangle,
\end{align}
where $s(z,\Delta,t)$ is the coefficient for the excitation in the ensemble, and $\phi(z,t)$ is the coefficient for the excitation of the light field. Using this general state in the Schr\"odinger equation \ref{eqn:schrodinger}the equations of motion for the state coefficients can be found:

\begin{align}
\frac{d}{dt}s(z, \Delta, t) &= -i \sqrt{2\pi}g^*  \phi(z,t) - i \Delta s(z, \Delta, t)\nonumber\\
\frac{d}{dt}\phi(z,t) &= -c\frac{d}{dz} \phi(z,t)- i\sqrt{2\pi} g \, \int d\Delta \, s(z,\Delta,t)\rho(z,\Delta).
\end{align}

The light travels through the system at a rate much faster than the evolution of the ensemble. This results in the time derivative of the light field becoming insignificant, giving the equations of motion:

\begin{align}
\frac{d}{dt}s(z, \Delta, t) &= -i \sqrt{2\pi}g^*  \phi(z,t) - i \Delta s(z, \Delta, t)\label{eqn:atoms}\\
\frac{d}{dz} \phi(z,t)&=- i\sqrt{2\pi}\frac{g}{c} \, \int d\Delta \, s(z,\Delta,t)\rho(z,\Delta)\label{eqn:light}.
\end{align}
These equations are solved with initial conditions $\phi(0,t) =0$, representing no light entering the ensemble, and $s(z,\Delta,0)$ found from the ASE solution using equation (\ref{Eqn:1photon}). 
The amplitude of the light field leaving the ensemble is given by $\phi(L,t)$. The chance of a photon being emitted between times $t_1$ and $t_2$ is given by $\int_{t_1}^{t_2}c|\phi(L,t)|^2 dt$, so the total probability of the ensemble emitting a RASE photon is given by

\begin{align}
P_\text{RASE} = \int_0^{\infty}c|\phi(L,t)|^2 dt,
\label{eq:probEmission}
\end{align}
where $0$ is the time at which the inverting pulse occurs between ASE and RASE.

\section{Effectiveness of RASE for single photon production}
\label{sec:analysis}

The effectiveness of RASE as a single photon source is judged by the efficiency of the emission of the RASE photon and the chance that there are multiple photons emitted close together. The spacing of photons is determined by the ensemble properties during the ASE process, while the efficiency of emission is calculated through the RASE simulations. In this section, we show that the RASE process can be made arbitrarily efficient at high optical depth, and the purity of the single photon state can be made arbitrarily high at low optical depth.

\subsection{Photon temporal distinguishability}
\label{sec:spacing}

To determine the purity of single photon detection, we must determine what temporal spacing is required in order to resolve separate photon emissions, and the probability of having more than one photon emitted into a particular mode within that timeframe.  We begin by considering emissions at a low coupling strength, where the equations can be investigated analytically. 

Detection of a single photon emission from an excited state ensemble leaves the ensemble in the state given by Eq.~(\ref{eq:singleEmission}), so the state coefficient $s(z,\Delta,t)$ in description of the general state (\ref{eqn:generalASE}) has the initial condition:
\begin{align}
s(z,\Delta,-T_S) = 1,
\end{align}
where $-T_S$ is the time that ASE photon is detected. Equation~(\ref{Eqn:1photon}) gives the evolution of this coefficient as a function of time. When the coupling of the atoms to the light is weak ($g^2 N/c << 1$) the latter two terms can be neglected and the equation of motion is

\begin{align}
\frac{d s(z,\Delta,t) }{dt} = \, &i \Delta \, s(z,\Delta,t).
\end{align}
The coefficient of the state of the system as a function of time is then
\begin{align}
s(z,\Delta,t) = e^{i \Delta (t-(-T_S))}.
\end{align}

We choose time $t = 0$ to be the time at which the ensemble state is inverted. This gives the initial state of the system after the inversion:

\begin{align}
|\psi_R(z,\Delta, 0)\rangle = \int dz \int d\Delta e^{i \Delta T_S} \hat \sigma_+(z,\Delta)|\{g\}\rangle.
\end{align}

Now we can calculate the re-emission profile of the light using equations (\ref{eqn:atoms}) and (\ref{eqn:light}). The RASE process is simplified when the coupling strength is low. In this regime the coupling of the atom excitation to light does not significantly change the atom excitation. The equations of motion can then be approximated as

\begin{align}
\frac{d}{dt}s(z, \Delta, t) &= -i \Delta s(z, \Delta, t)\\
\frac{d}{dz} \phi(z,t)&=- i\sqrt{2\pi}\frac{g}{c} \, \int d\Delta \, s(z,\Delta,t)\rho(z,\Delta).
\end{align}

The atom excitation equation can be then solved explicitly, with initial condition $s(z,\Delta,0) = \rho(z,\Delta) e^{-i \Delta (T_S)}$ to give

\begin{align}
s(z,\Delta, t) = \rho(z,\Delta) e^{-i \Delta (t-T_S)}
\end{align}
We can see that when $t = T_S$, all the components of the atomic excitation are in phase. Solving the light coefficient equation gives the light emitted from the ensemble

\begin{align}
\phi(L,t) &= - i\sqrt{2\pi}\frac{g}{c} \int dz \int d\Delta \rho(z,\Delta) e^{i \Delta (T_S+t)}\\
&\propto \int d\Delta \rho(z,\Delta) e^{i \Delta (T_S+t)}.
\end{align}

This is the Fourier transform of the ensemble spectral density. For a Gaussian distribution with width $\Gamma$, for example, this is also Gaussian, with width $\tau = 1/\Gamma$.  A spacing of adjacent emissions of $4\tau$ will result in negligible overlap ($< 5\%$) between their photon profiles. For other distributions of ensemble inhomogeneity there will be similar spacing conditions.

Given this criterion for resolving single photons, we now examine the likelihood that the source will only emit a single photon within the $4\tau$ timeframe.  This is trivially true for arbitrarily low coupling strength, where the photon emission rate goes to zero.  At high coupling strengths, we must solve the equations of motion numerically. 

The collective coupling strength of the ensemble is best characterised by the optical depth ($\alpha L$): a dimensionless quantity, which for a uniform ensemble is equal to the absorption coefficient $\alpha$, when the ensemble is in the ground state, multiplied by the length $L$ of the ensemble. It defines how much light can be transmitted through the ensemble, but also encapsulates how much spontaneous emission from the ensemble will be amplified through stimulated emission. Using an ensemble with a Gaussian distribution in detuning with width $\Gamma$ gives a peak optical depth of $\alpha L = \frac{2\pi g^2}{c \Gamma} N$, where $g$ is the coupling strength of each individual atom to the vacuum, $c$ is the speed of light, and $N$ is the number of atoms in the ensemble.

\begin{figure}[h]
\begin{center}
\includegraphics[width=30pc]{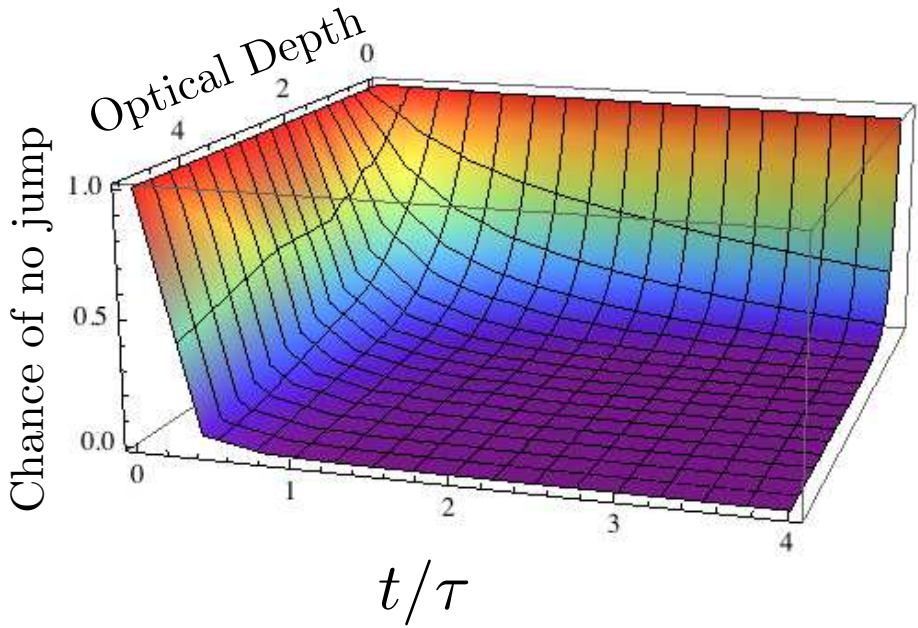}
\end{center}
\caption{\label{fig:optdScan3D} Probability of having no second emission after the first emission as a function of time and optical depth. For the RASE single photon source to be effective the ASE emissions have to be well-spaced, with a spacing of $4\tau$ giving an overlap of two emissions to be less than $5\%$. These simulations show that for optical depths of greater than $\alpha L = 0.2$, consecutive emissions are more than 50\% likely to be closer than $4\tau$ apart.
}
\end{figure}

We solve equation (\ref{Eqn:1photon}) numerically to simulate an ensemble with the Gaussian spectral distribution above, and calculate the probability (\ref{eq:noJump}) that no second photon is emitted within the $4\tau$ resolution window for a range of optical depths.  The results are shown in figure \ref{fig:optdScan3D}, where we see that for an optical depth greater than $\alpha L = 0.2$, consecutive RASE emissions are likely to overlap.

\subsection{Photon rephasing efficiency}

We have seen above that single photon purity is best at low optical depth, where there is sufficiently low rate of photons emissions that it becomes highly unlikely that two photons will be emitted within the temporal width of the photon.  Unfortunately, the exact same reasoning leads us to expect that when the excitation in the inverted ensemble rephases, it is concomitantly unlikely to emit the photon.

\begin{figure}[htb]
\begin{center}
\includegraphics[width=30pc]{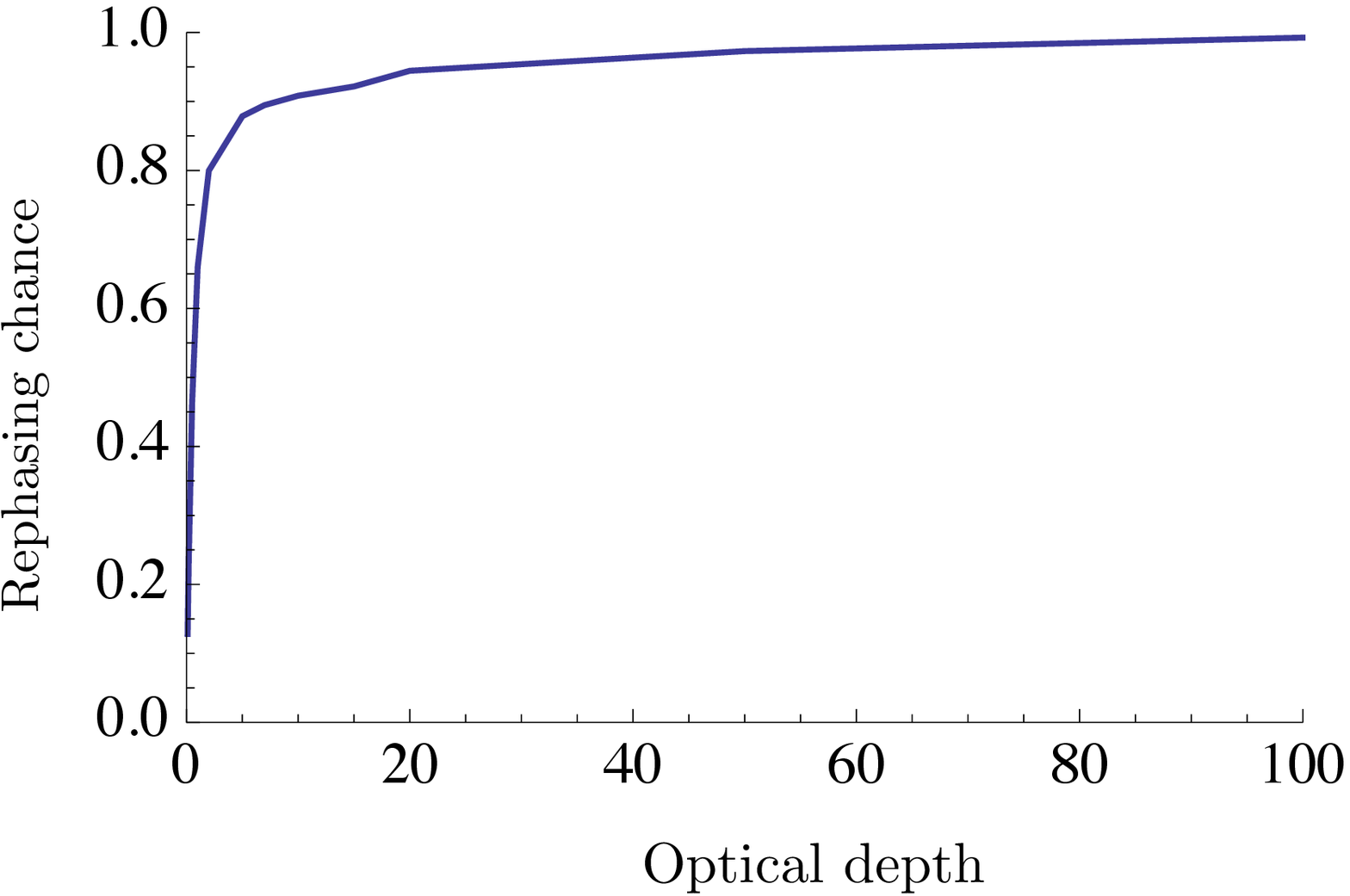}
\end{center}
\caption{\label{fig:CondRase} Probability of a photon being rephased during RASE, as a function of optical depth. The simulations here assume a single photon has been emitted from an atomic ensemble in the excited state, then $4\tau$ time elapses without further emissions, and then the ensemble is inverted. The chance that the ensemble re-emits the photon is calculated to be the integrated intensity of the light field leaving the ensemble at $z = L$.}
\end{figure}

In figure~\ref{fig:CondRase}, we calculate this probability of re-emission for a range of optical depths, using equation (\ref{eq:probEmission}). 

\begin{figure}[htb]
\begin{center}
\includegraphics[width=30pc]{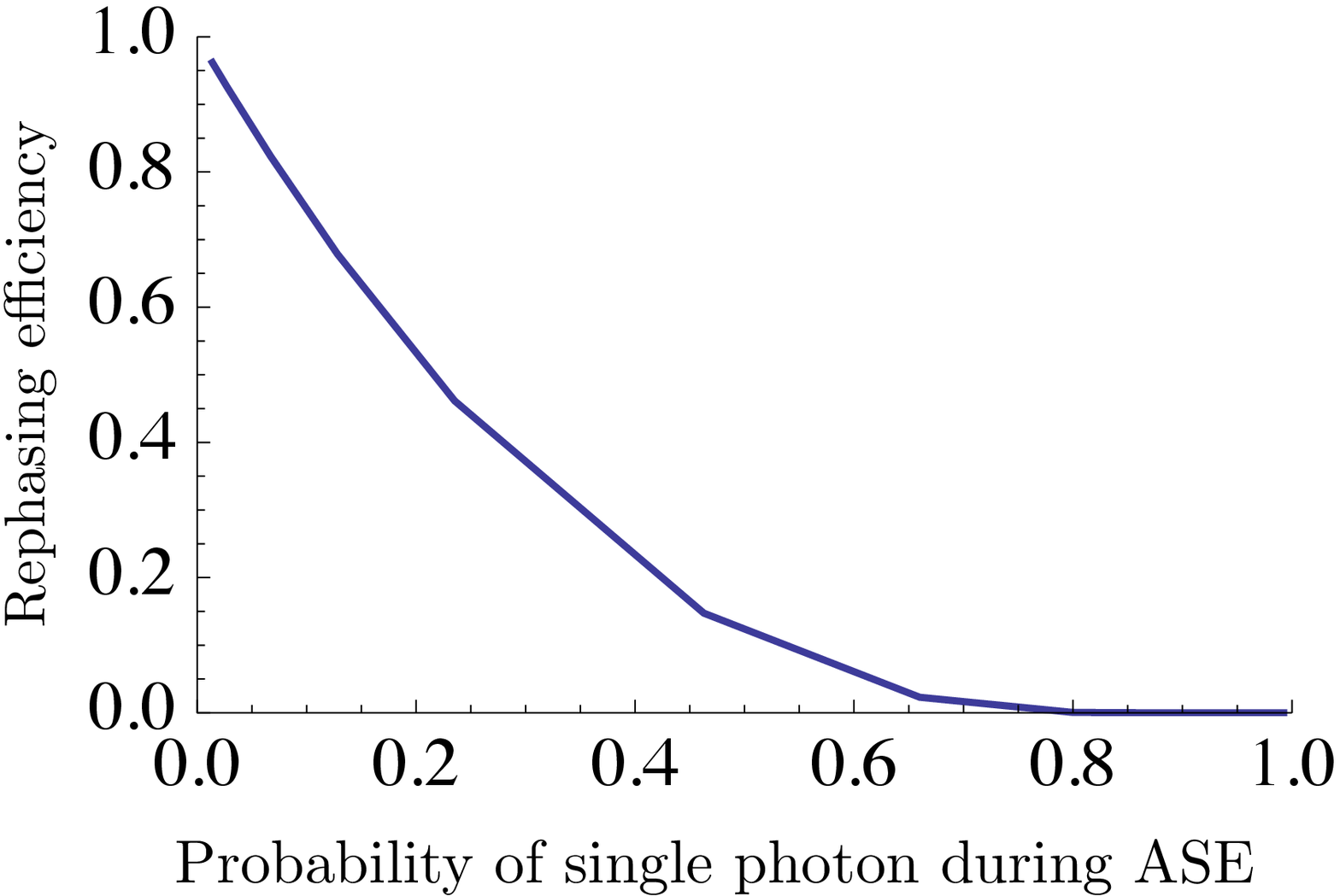}
\end{center}
\caption{\label{fig:chancecomparison}
A comparison of the separation of photons during ASE and the efficiency of emission during RASE at different optical depths. The x-axis is the chance that adjacent photon emitted during ASE are separated by at least $4\tau$, and the y-axis is the chance that a RASE photon is emitted. At low optical depths the ASE photons have a high chance of being well separated, but have a low chance of being rephased. At high optical depths the efficiency of rephasing approaches 1, but the chance of having only a single photon approaches 0.
}
\end{figure}

To quantify the trade-off, we plot the efficiency against the purity of the single photon source in figure~\ref{fig:chancecomparison}. 

In Section~\ref{sec:tailored} we investigate how this trade-off can be avoided, and we can ensure that the source produces well-separated photons at high efficiency.

\section{Improving the efficiency of the single photon source}
\label{sec:tailored}
As it stands, the RASE protocol described in section~\ref{sec:model} is not a particularly efficient source of resolvable single photons. For photons that are 60\% likely to be separated from their neighbours, the efficiency of rephasing the echo photon is limited to less than 10\% (see Fig.~\ref{fig:chancecomparison}). With sufficiently efficient number-resolving detectors this problem can be partially overcome by observing the ASE emissions collected by the photodetector and only selecting out photons that are resolvable from their neighbours. Current photodetectors are not efficient enough to guarantee with high fidelity that only a single photon was emitted. Additionally, at high optical depth the probability of having well-separated photons is too low for this operating to be an efficient sources of single photons. In this section we investigate a method of reducing the chance of photons being emitted close together while increasing the rephasing efficiency.

\subsection{Different optical depths for emission and rephasing processes}

Given that the emission and rephasing processes work best in completely different optical depth regimes, it would make sense to try to perform each stage at the most favourable region of parameters. The problem with that is that the optical depth is fixed by the properties of the ensemble and the two-level atomic transition considered.

A four-level photon echo~\cite{Beavan:2011} provides a solution to this problem. In a four-level photon echo the ASE and RASE photons are produced using different transitions. This allows for different vacuum coupling strengths during the ASE and RASE periods. The ASE photon can be emitted from a low coupling strength transition, which increases the separation between adjacent photons, and the RASE photon from a high coupling strength transition, increasing the emission efficiency. Although there are four levels involved in the scheme, there is only coherence between two at a time, so the modelling used here still applies.

\begin{figure}[ht]
\begin{center}
\includegraphics[width=30pc]{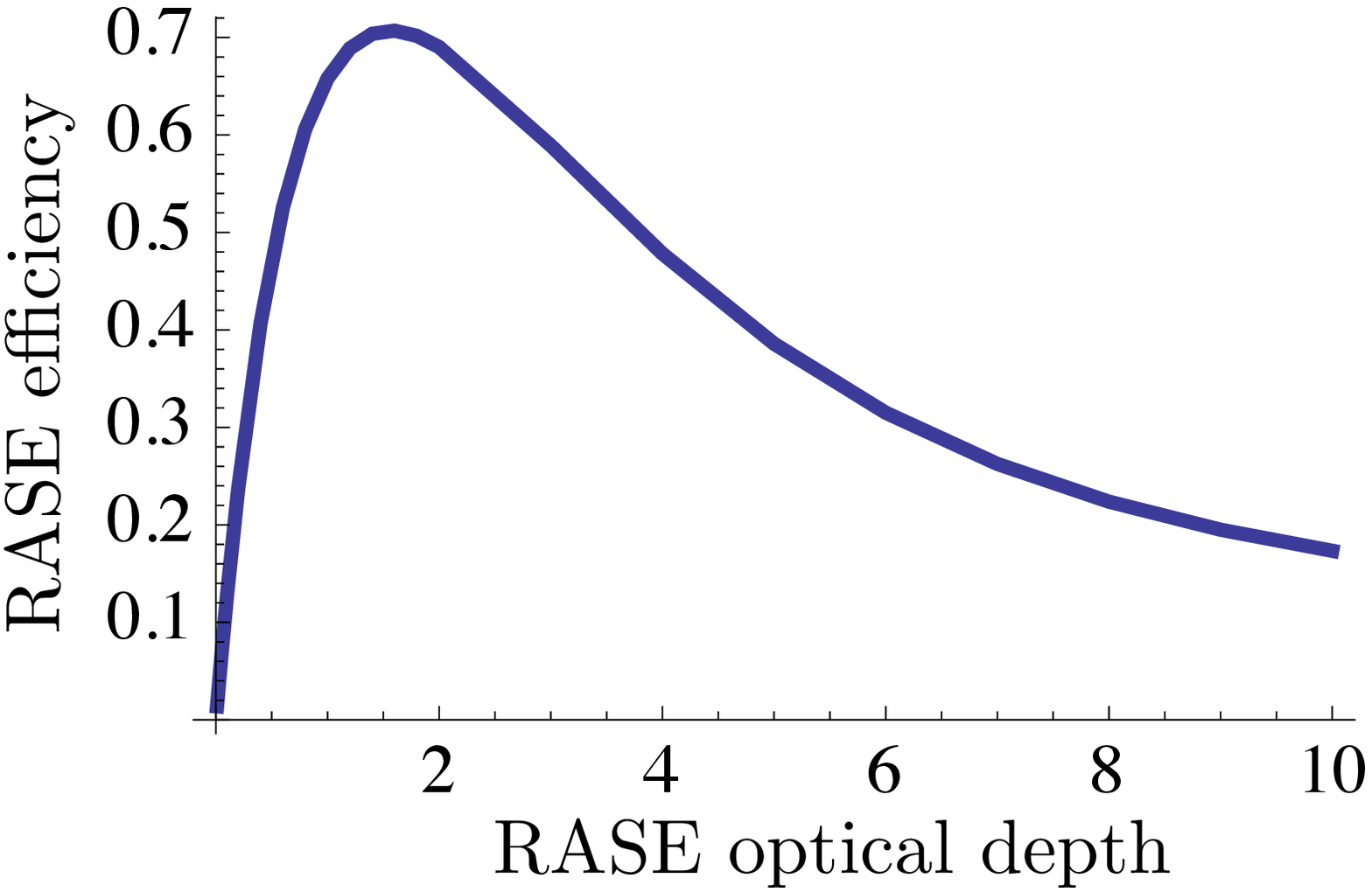}
\end{center}
\caption{\label{fig:LightOutFlat} Probability of a RASE photon being emitted as a function of optical depth, when the ASE preparation happens at low optical depth. At low optical depth the atoms emit independently, so the state of the ensemble after an ASE emission is an even superposition of each atom emitting a photon. The RASE efficiency in this situation peaks near an optical depth of 1, at around 70\% efficiency, and this efficiency decreases as the optical depth is increased past this point.
}
\end{figure}

For the ASE emission, we have chosen a probability greater than 95\% that the overlap of neighbouring photons is less than 5\%. This corresponds to a peak optical depth of less than 0.1. For optical depths this low, the collective de-excitation is evenly distributed among the atoms, as shown in section~\ref{sec:distribution}. To find the optimal optical depth for the RASE process we used this even distribution as the initial state for simulations of RASE with higher coupling strengths. The efficiency of RASE as a function of coupling strength is plotted in figure~\ref{fig:LightOutFlat}.

\begin{figure}[htb]
\begin{center}
\includegraphics[width=30pc]{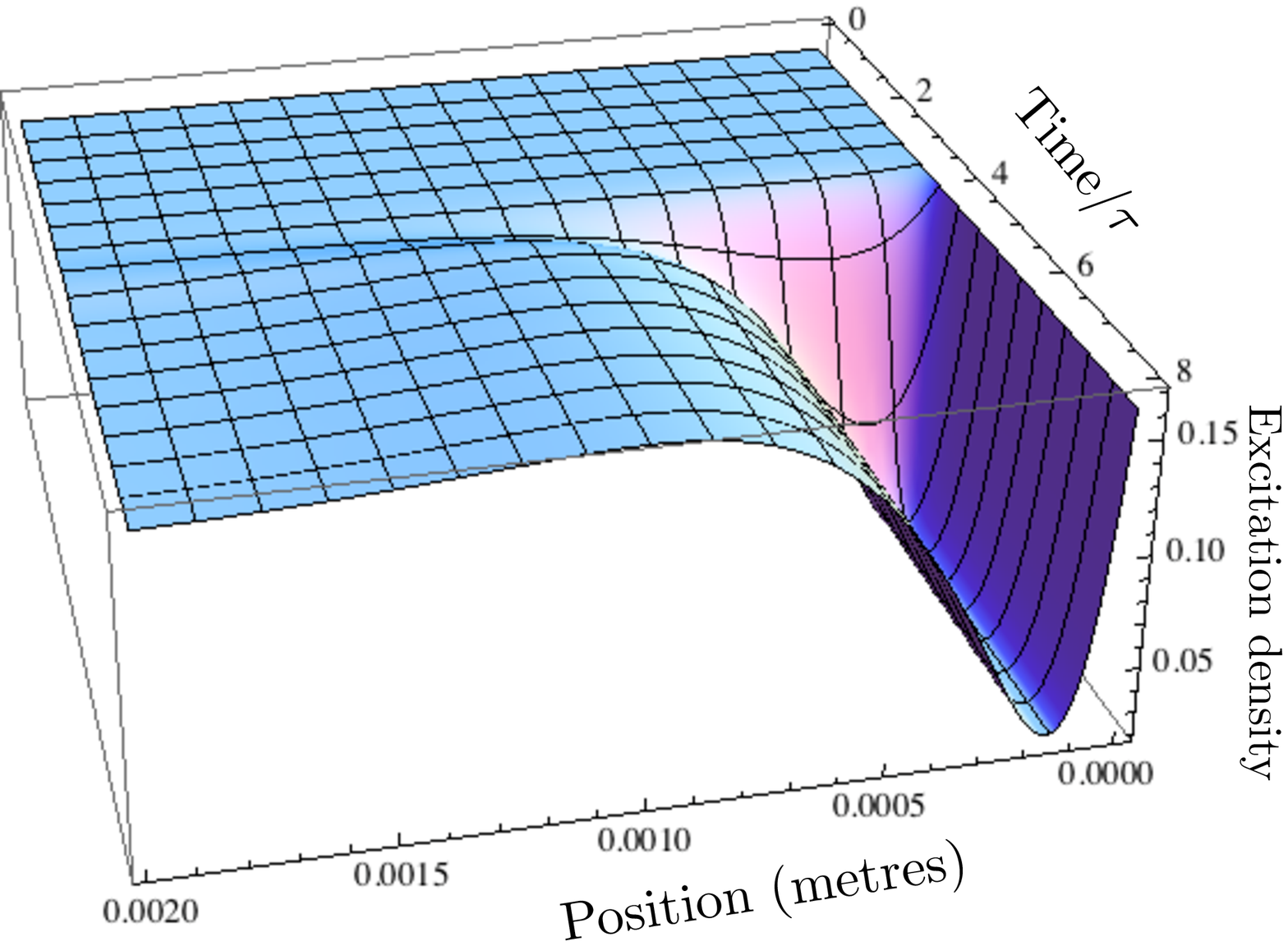}
\end{center}
\caption{\label{fig:HighDepth} Spatial distribution of excitations in the atomic ensemble as a function of time, at an optical depth of 10. The emission exits the ensemble to the left of the figure, and time runs from the back of the figure to the front. The initial condition is a de-excitation happening $4\tau$ before a $\pi$-pulse inverts the excitation, so the re-emission is centred at $4\tau$ after the $\pi$-pulse (applied at $t=0$).
}
\end{figure}

These simulations show that the probability of rephasing a photon does not increase arbitrarily with the optical depth, instead peaking near $\alpha L = 1$ with a probability of rephasing of approximately 70\%. The efficiency is limited by reabsorption of light emitted from the back of the ensemble before it has travelled through to the front. Figure~\ref{fig:HighDepth} shows the spatial profile of the collective excitation for an ensemble with an optical depth $\alpha L = 10$. The atoms at the back of the ensemble de-excite almost fully, but the light emitted from these atoms is absorbed further down the ensemble.

\subsection{Tailoring the ensemble spectral density}

Even though the strategy of using different couplings to the ASE and RASE stages allowed us to obtain mostly pure single photon states, it had limited efficiency due to reabsorption in the ensemble. To understand our solution to this problem we should go back to the results from Fig.~\ref{fig:CondRase}. There, when the ensemble was prepared assuming a single excitation at high optical depth, the efficiency of RASE approached 1. The main difference between the two cases lies in the de-excitation profile imprinted in the ensemble by the ASE process. While at low optical depths the profile is flat, for high optical depths it has a strong concentration at the front of the sample. 

The results from Fig.~\ref{fig:CondRase} were obtained from this initial highly asymmetric ASE profile, which is mode-matched to the emission of a RASE photon at the same optical depth. This led to unity efficiency at the cost of having very low probability of producing a single isolated photon. The strategy that we propose when the optical depth is changing between ASE and RASE, is to create this mode-matching artificially. For this, one needs to tailor the density profile of the ensemble to match the required excitation distribution.

\begin{figure}[htb]
\begin{center}
\includegraphics[width=30pc]{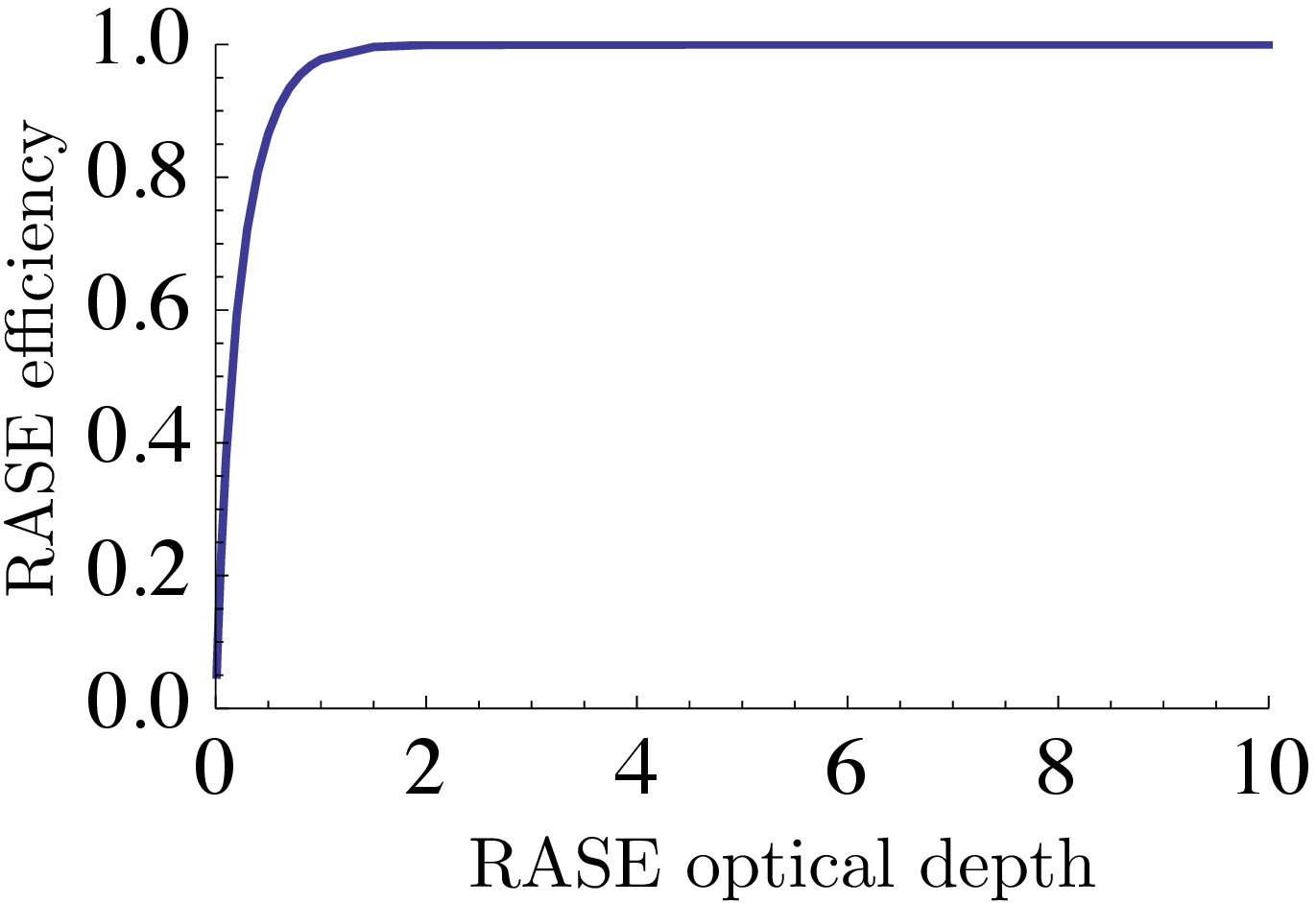}
\end{center}
\caption{\label{Fig:Tailored} Efficiency of photon rephasing with RASE using a tailored density profile. The shape of the density was found by simulating an emission of a single photon from an ensemble with a flat density in position and a Gaussian density distribution in detuning, with chosen optical depth. The distribution of the de-excitation left by that collective emission was used as the shape of the density $\rho(z,\Delta)$ for ASE of a single photon at low optical depths, where every atom emits independently. The ensemble is then inverted and the optical depth changed to match the optical depth used to find the density, and the efficiency of the RASE photon emission is measured. This emission was found to be more efficient than those in section \ref{sec:analysis} and in addition is not restricted by optical depth during ASE, as a peak optical depth lower than 0.1 ensures a high chance of well-separated photons.
}
\end{figure}

In this new scenario, the emission will be equally distributed among all the active atoms in the ensemble (due to the low optical depth condition), but there will be more atoms in the regions that require a larger share of the excitation. Figure~\ref{Fig:Tailored} shows the rephasing efficiency for these tailored density profiles. For each optical depth in this figure, the required density profile $\rho(z,\Delta)$ was found by simulating ASE at that optical depth. The initial condition for the RASE simulations was then found by simulating ASE from this tailored density profile at a low optical depth, resulting in an equal superposition of all active atoms in the ensemble. The evolution of the RASE process was then calculated as a function of optical depth. These simulations show that the reabsorption can be completely eliminated, while still operating ASE at a low optical depth to produce well-separated photons.

\section{Conclusions}

In conclusion, we have shown that RASE can be used as an efficient and high fidelity single photon source. The use of different optical depths for the emission and rephasing stages, combined with a careful shaping of the atomic density in the ensemble, allows photons to be produced sufficiently far apart to be treated as single, while allowing their emission probability to be high. 

Implementing these strategies is well within currently available technology. The use of different atom-light coupling strengths can be achieved by using the four-level photon echo scheme presented in~\cite{Beavan:2011}, while tailoring the atomic density can be obtained by targeted spectral hole-burning technique. 

Our analysis was done under idealised conditions and the next step towards a realistic implementation would be to consider various experimental imperfections. An important factor in this system is the multi-level structure used to provide the different coupling strengths. This introduces extra incoherent decay channels~\cite{Beavan:2012, Hedges:2010} that would increase the noise on the output and degrade the single photon production.

Current experiments have put the ensemble in a cavity to enhance the coherent transition. This introduces different conditioned dynamics on the ensemble, and research is ongoing in this direction. Once single photon production using RASE is successful, this scheme can produce spatially separated entangled ensembles that can be used for entanglement swapping as part of a quantum repeater. If the outputs of two ensembles are combined on a beamsplitter, detection of a photon projects the two ensembles in a entangled state. In this setup, the ensemble acts as both the single photon source and the quantum memory required for a quantum repeater, increasing the efficiency of the whole system.

\ack

MJS and ARRC gratefully acknowledge support by the Australian Research Council Centre of Excellence  for Quantum Computation and Communication Technology (Project No. CE110001027). MJS also acknowledges the support of DSTO. JJH was supported by an Australian Research Council Future Fellowship (FT120100291).

\bibliographystyle{unsrt}
\bibliography{Papers}

\end{document}